\documentclass[fleqn,twoside]{article}
\usepackage{espcrc2}

\usepackage{graphicx}
\usepackage[figuresright]{rotating}

\newcommand{\AmS}{{\protect\the\textfont2
  A\kern-.1667em\lower.5ex\hbox{M}\kern-.125emS}}
\def\neut{\tilde{\chi}^0_1}
\def\lsp{\tilde{\chi}^0_1}
\def\neutt{\tilde{\chi}^0_2}

\def\stau{\tilde{\tau}_1}

\def\et{E_T^{\rm miss}}
\title{Dark matter and the LHC}

\author{G. B\'elanger\address[MCSD]{LAPTH, Universit\'e de Savoie, CNRS \\ 
       B.P.110, F-74941 Annecy-le-Vieux, France}%
       }

\begin{document}

\begin{abstract}
Cosmological and astrophysical measurements indicate that the universe contains a large amount
of dark matter. A number of weak scale dark matter candidates have been proposed in extensions of the
standard model. 
The potential to discover the dark matter particle and determine its properties at the upcoming LHC 
 is summarized.
\vspace{1pc}
\end{abstract}

\maketitle

\section{INTRODUCTION}

There is strong evidence coming from various scales that dark matter (DM) dominates over luminous matter in the universe. 
In recent years the relic abundance of DM
has been extracted with good precision from cosmological measurements, $\Omega h^2=0.1099\pm 0.0062
$~\cite{Dunkley:2008ie}.
A favoured explanation for this DM is  a new neutral and stable weakly  interacting particle. 
A variety of candidates in extensions of the standard model have been proposed~\cite{Bertone:2004pz}.
The best motivated candidates are those that arise in models constructed to solve the electroweak 
symmetry breaking problem.
This includes: the Majorana neutralino in either the minimal supersymmetric standard model (MSSM), its  
singlet extensions such as the NMSSM or nMSSM~\cite{Stephan:1997ds} or in  GUT-scale
supersymmetric models;
the right-handed Dirac neutrino in models of warped extra dimensions or in models with left-right symmetry~\cite{Agashe:2004ci};
the gauge bosons or scalar photons in  universal extra dimension models~\cite{Appelquist:2000nn,Dobrescu:2007ec};
the gauge bosons of the little Higgs model~\cite{Cheng:2003ju}; the 
right-handed sneutrino of supersymmetric models~\cite{ArkaniHamed:2000bq}; the   scalar
 in extensions of the standard model~\cite{McDonald:1993ex}. 
To this extensive list one should add new phenomenologically 
inspired DM candidates~\cite{strumia} as well as candidates with super weak interactions such as the gravitino~\cite{Buchmuller:2003is}.

The single measurement of the relic abundance, although precise, is 
not enough to elucidate the nature of dark matter and distinguish the various candidates.
For this, searches in astroparticle and
 collider experiments are being pursued actively.
Direct detection experiments would unambiguously establish that a stable particle constitute 
the DM. For now the upper bounds on the elastic scattering cross section constrain the DM models, 
although some dependence on astroparticle quantities, such as  the DM distribution are introduced.
The indirect detection rates also depend on the DM annihilation cross section,  however astrophysical 
sources could lead to signals that might be hard to distinguish from those of DM.
For example pulsars provide a possible explanation to the latest positron excess observed by PAMELA~\cite{Adriani:2008zr}.

This leaves a double challenge for dark matter studies at colliders and in particular at the LHC. 
The first goal is the search for the DM candidate and/or other new 
particles predicted in the framework of the various theoretical models. 
The second, if a signal is found, would be to determine the properties of the DM  
particles, its mass, spin and couplings. This information  could then be used to  
infer the DM annihilation and elastic scattering cross sections.
A comparison with the value of 
 the relic abundance extracted from cosmological observations would then allow to test the underlying cosmological 
model. For example the relic density can be reduced by orders of magnitude  in
non-standard scenarios with low reheat temperature and/or late entropy production~\cite{Gelmini:here,Drees:2007kk}. 
A comparison with (in-)direct detection rates would allow self-consistency checks and  provide additional  information on 
quantities such as the DM distribution, the propagation models, etc...
How well  the properties of DM can be determined strongly depend on the particle physics model. Below we summarize the main
results for the LHC discovery potential as well as for parameter determination in three extensions of the
standard model: 
the MSSM, the universal extra dimensions model and a little Higgs model.

\section{NEUTRALINO  IN SUPERSYMMETRY}

In supersymmetric models, DM candidates include the neutralino, the partner of the gauge and Higgs bosons,
as well as the gravitino~\cite{Buchmuller:2003is} and the axino~\cite{Covi:1999ty}. 
We will here concentrate on the weakly interacting  candidate, the neutralino,
which has the richer collider phenomenology as well as the best prospects for (in-)direct detection. 
The neutralino DM and the collider SUSY phenomenology were
first analysed in the context of a constrained model with a small and manageable number of arbitrary parameters 
defined at the GUT scale, the CMSSM. The model parameters are the common scalar mass, $m_0$,
the common gaugino mass, $m_{1/2}$, the trilinear coupling, $A_0$ and the ratio of the vevs of the two Higgs doublets 
$\tan\beta$ as well as the sign of $\mu$. It has since been realised that properties of DM might not be 
confined to the preferred bino scenario of the CMSSM  thus expanding the range of possibilities for DM searches in various  
GUT scale motivated models as well as in extensions of the MSSM.

At the LHC, a pp collider of 14TeV, since the direct production of DM does not leave a good signature, 
the  best channels to detect  DM particles are  via the
production and decay of heavy coloured particles. The signature of the DM candidate always  produced in the decay chains 
involves observables such as missing $E_T$ or $p_T$.

Supersymmetric models contain new coloured particles, the squarks and gluinos
 partners of the standard quarks and gluons. 
The cross sections for production of these particles at a hadronic collider are large, 
they can reach $10^2$ pb's for sparticles of 500~GeV.  
Production of quarks or gauge bosons in SM processes is however orders of magnitude larger, finding   
efficient ways to cut the background and enhance the signal is therefore a critical issue. 
Furthermore in SUSY models electroweak production of new particles, in particular the
charginos and neutralinos partners of the gauge and Higgs bosons can occur. These constitute the
 dominant production process in scenarios
where the coloured spectrum is heavy.  

\begin{figure}[htb]
\vspace{9pt}
\includegraphics[width=7cm]{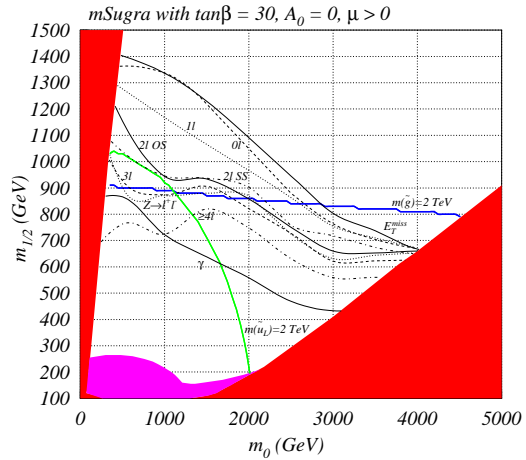}
\vspace{-0.8cm}
\caption{The LHC reach for SUSY in the CMSSM with 100 fb$^{-1}$. The signal is observable below the contour corresponding
to each event topology, from ~\cite{Baer:2008uu}.}
\label{fig:LHCreach}
\end{figure}

There are a variety of  SUSY signatures at the  LHC. To estimate the reach of the collider it is
convenient to classify the signatures according to the number of leptons (of same sign or opposite sign)
accompanying the  $\et$. 
The reach for each event topology  at the LHC with luminosity
of ${\cal L}=100$~fb$^{-1}$ is displayed in Fig.~\ref{fig:LHCreach} in the $m_0-m_{1/2}$ plane of the  CMSSM.
The reach is almost 2TeV for gluinos in scenarios 
where squarks are very heavy (large $m_0$) and can otherwise reach  3TeV for squarks and gluinos~\cite{Baer:2008uu}. 

\begin{figure}[htb]
\vspace{9pt}
\vspace{-1.2cm}
\includegraphics[width=8.5cm]{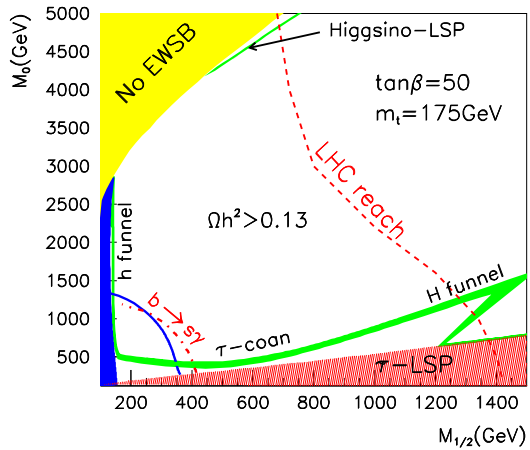}
\vspace{-2.2cm}
\caption{The regions where $\Omega h^2=0.11\pm0.02$ (green) in the CMSSM with $\tan\beta=50$ and 
the exclusions from  $m_h>114$~GeV and 
$b\rightarrow  s\gamma$, from~\cite{Belanger:2004ag}. }
\label{fig:wmap}
\end{figure}

Within the CMSSM it means that the LHC could discover SUSY in different channels in most of the region where $\Omega h^2<0.11$. 
This includes, see Fig.~\ref{fig:wmap} 
\begin{itemize}
\item{} The  region at low $m_0-m_{1/2}$ where a bino LSP annihilates into light fermions through sfermion exchange. 
A fraction of this region is however constrained by Higgs searches and flavour physics.
\item{} Part of the sfermion coannihilation region at low values of $m_0$.
\item{} The region at low $m_{1/2}$ where the bino LSP annihilates through a light Higgs resonance
\item{} Part of the region where the LSP annihilates through a heavy Higgs resonance    
at large values of $\tan\beta$.
\item{} Part of the 'focus point' region where the LSP is a  bino/Higggsino LSP. This region is associated
with heavy squarks ($m_0>3TeV$) and only the gluino is light enough to be produced. Direct production of
neutralino/chargino can also be used for SUSY discovery. 
\end{itemize}

Note that when allowing all free SM and CMSSM parameters to vary, comprehensive  fits to available data 
show that the thin strips displayed in Fig.2 extend to 
wide regions of the $m_0-m_{1/2}$ plane~\cite{roskowski,Allanach:2006cc}.

The CMSSM might be a much too contrived scenario. Allowing additional parrameters in GUT scale model, for example non
universality of the scalars or the gauginos,  or even going to the full MSSM model with parameters defined at the weak scale
will open up the possibilities for DM annihilation.  The LHC discovery potential in terms of 
squark and gluino masses should not be much affected if sparticle masses are not degenerate.  The new DM scenarios include
\begin{itemize}
\item{} The bino/wino LSP which annihilates into gauge boson pairs~\cite{Baer:2005zc}.
\item{} The bino/Higgsino LSP associated with TeV scale sfermions.
\end{itemize}

\subsection{Determination of particle properties}

For reconstructing the DM annihilation cross sections the quantities that need to be measured
are the mass and couplings of the LSP, the mass of new particles  (or  lower limits)
that contribute to DM 
(co-)annihilation and the mass of  any  resonance that can enter the LSP annihilation.

The difficulty in determining  parameters  of the DM model at the  LHC is the large amount of missing energy 
that prevents the reconstruction of a mass peak. The standard method for mass determination relies rather 
on measuring end points in kinematic distributions~\cite{Bachacou:1999zb}. Consider    
the standard decay of a squark, 
$\tilde{q}\rightarrow q\neutt\rightarrow q l^\pm\tilde{l}^\mp\rightarrow
ql^\pm l^\mp \neut$ with a signature in lepton pairs, quarks and $\et$. 
The upper edge of the $ll$ distribution
and both edges of the $llq$ and $lq$ distributions depend on the mass differences of the 
sparticles occurring in the decay chain, for example
\begin{equation}
m_{ll}^{max}=\sqrt\frac{(m^2_{\neutt}-m^2_{\tilde{e}_R})(m^2_{\tilde{e}_R}-m^2_{\neut})}{m^2_{\tilde{e}_R}}.
\end{equation}  Case 
studies have shown that  the precision achievable on the mass differences
is typically at the few per-cent level ~\cite{Bachacou:1999zb}.  This  assumes  
that the particles in a decay chain are
correctly identified which might require making some assumptions on the theoretical model.
Combining this method with cross sections measurements can improve the parameter determination
 ~\cite{Lester:2005je}.

The precision that could be achieved  on a collider  prediction of the relic abundance,
$\Delta \Omega/\Omega$, was studied in a few generic scenario and benchmark analyses were performed for
both ATLAS and CMS. 

In  CMSSM scenarios where the LSP is a light bino, annihilation into fermion pairs through sfermion
exchange  dominate together with bino/stau coannihilation. 
The relevant parameters are the LSP mass, the LSP couplings and the slepton masses. 
It was shown that to have a 10\% precision on a collider prediction of the relic abundance would require
a precise measurement of  the mass difference $\Delta(m_{\stau}-m_{\lsp})$  (at the \%
level) while other parameters of the neutralino sector ($M_1,\mu,\tan\beta$) need to be measured at the 10\% level~\cite{Allanach:2004xn}. 
A study of a benchmark point (SPS1a') which belongs to this 
class of scenario was performed by two groups for the LHC. In ~\cite{Nojiri:2005ph}, the endpoint methods mentionned above 
was used  to reconstruct the four masses $\neut,\neutt, \tilde {l},\tilde{q}$  in squark decays.
The complete determination of the neutralino sector also requires  measuring  the chargino mass
and extracting $\tan\beta$ from the  Higgs sector.  Furthermore the mixing in the stau sector which
also inflences the neutralino couplings is determined from the leptonic branching fractions $Br(\neutt\rightarrow
\tilde{l}_R l)/Br(\neutt\rightarrow \tilde{\tau}_1 \tau)$. The precision achievable 
with ${\cal L}=300{\rm fb}^{-1}$ is $\Delta\Omega/\Omega \approx
20\%$. Similar results were obtained in~\cite{Baltz:2006fm}.
Improving the determination of the neutralino and stau masses as well as the neutralino ouplings as could be done at the ILC 
would drastically reduce the uncertainty in the $\Omega h^2$ prediction to the
few percent level~\cite{Baltz:2006fm}. For these scenarios the collider
prediction for the elastic scattering cross section are expected to have 
almost an order of magnitude uncertainty.

\begin{figure}[htb]
\vspace{9pt}
\includegraphics[width=7cm]{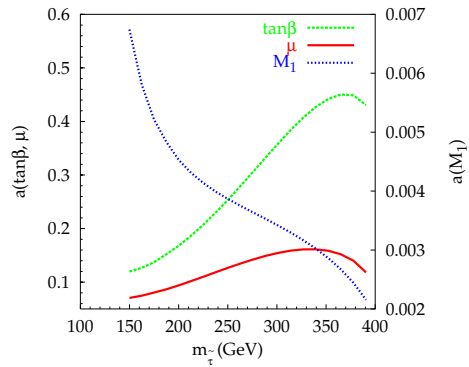}
\vspace{-0.8cm}
\caption{Required precision  on $\tan\beta,\mu$ (left axis) and $M_1$ (right axis) 
for  a 10\% precision on the collider prediction of $\Omega h^2$  in  CMSSM bino scenarios, from ~\cite{Allanach:2004xn}.}
\label{fig:delta}
\end{figure}

An even  more challenging scenario is the one where the LSP is a mixture of bino and Higgsino. 
In the CMSSM this means that scalars are very heavy while in the general MSSM such 
scenarios can occur also for a (sub-)TeV scale sfermion sector. The sfermion sector is not needed for
DM annihilation aince the self-annihilation of 
the bino/Higgsino into W pairs through chargino or boson exchange is efficient enough.
The annihilation is governed by the bino/Higgsino fraction of the LSP which  in turn  depends strongly on
$M_1$ and $\mu$. To make a 10\% prediction of the relic abundance
would require  to know these two parameters at the 1\% level. This level of precision is a real 
challenge for the LHC. In this scenario the gluinos are the only coloured particle that can be produced. 
The gluinos decay into quarks and heavy neutralinos/charginos which in turn decay into leptons
and the LSP. In a 
benchmark study done for ATLAS~\cite{DeSanctis:2007td},
 it was shown that  exploiting end point measurements the gluino mass as well as 
the  mass difference between neutralinos and the LSP could be determined with a  10\% accuracy.
This clearly leaves a large uncertainty in  the prediction of $\Omega h^2$, roughly a factor 4 ~\cite{DeSanctis:2007td}.

Finally we comment on the importance of knowing the complete Higgs spectrum. 
Consider for example a simple extension of the MSSM with one additional singlet superfield, the NMSSM. 
This model contains an extra neutralino as well as new scalars. Because of 
efficient neutralino annihilation through a scalar resonance, one finds that $\Omega h^2\approx 0.1$
in regions of parameter space where one expects $\Omega h^2\approx 1$ in the CMSSM. 
After the  measurement of SUSY parameters at the LHC one would likely conclude that the collider relic abundance 
does not match the cosmological measurement. It would be difficult  
to differentiate the non-standard cosmological scenario from a new particle scenario  unless the new scalar is
discovered and that is challenging~\cite{Belanger:2008nt}.

\section{UNIVERSAL EXTRA DIMENSION MODEL}

Models with exta dimensions offer an alternative solution to the hierarchy problem. 
In the minimal universal extra dimension model (MUED) all fields propagate in the bulk.
Each SM particle is accompanied by a tower of KK particles, the new particles have the same spin as the standard ones. The DM candidates are either  the first KK level of the
 hypercharge gauge boson, $B^1$ or the KK graviton.
We will consider only the former possibility since the graviton  has small detection rates.  
The DM being a gauge boson  typically annihilates more efficiently than the neutralino in supersymmetry with
important  annihilation  channels  into  $f\bar{f}$. A relic abundance within the range determined by WMAP requires,
assuming the standard cosmological scenario, a rather heavy $B^1$ in the 500-900GeV range~\cite{Kakizaki:2006dz,Kong:2005hn}. A lighter DM implies 
a low relic abundance. One feature of this model is that the annihilation into light fermions
in particular $e^+e^-$ is not suppressed, favouring a large indirect detection rate for positrons. 
A typical signal at LHC is the production of heavy quarks decaying into a gauge boson and a quark, 
the leptonic decay of the neutral gauge boson leads to a signature into  4 leptons and $\et$.
In this channel the LHC discovery reach is the TeV scale ~\cite{Cheng:2002ab}.
The  cascade decays of KK or SUSY particles are very similar, one difference
is that the KK particles are almost mass degenerate so that the outgoing fermions in the cascade decays can be very soft. 
More importantly the spin of the new particles differ by an half-integer in the two models.
Different asymetries have been proposed for spin identification at colliders~\cite{Barr:2005dz}.
An explicit example on  how to use angular correlations to differentiate the fermionic gluino in SUSY from the
KK gluon was presented in~\cite{Alves:2006df}.
Note that the production of new resonances, for example a  $Z^1$ decaying into fermions offers a distinctive  
signature of this model as compared to SUSY.

\begin{figure}[htb]
\vspace{9pt}
\includegraphics[width=7cm]{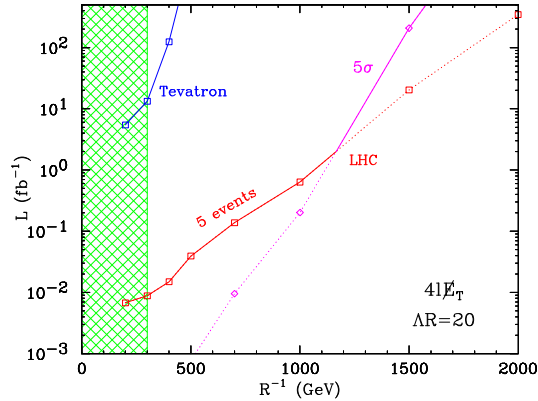}
\vspace{-0.8cm}
\caption{Luminosity required at LHC for a 5$\sigma$ signal in the $4l+\et$ channel in the UED model, 
from ~\cite{Cheng:2002ab}.}
\label{fig:ued}
\end{figure}

\section{LITTLE HIGGS MODEL}
As a final example, consider the little Higgs model where the Higgs is a pseudo Goldstone boson 
originating from the spontaneous breaking of a global symmetry at a higher scale. The global symmetry 
protects the Higgs mass from large corrections.
Strong  constraints from electroweak precision observables can be
avoided by imposing  T-parity.  The model contains new heavy gauge bosons as well as
heavy top quarks (both T-odd, $T_-$  and T-even, $T_+$).   The DM candidate is the lightest new heavy neutral
gauge boson $A_H$  that annihilates preferentially 
through Higgs exchange into W pairs~\cite{Cheng:2003ju}. The minimal version of this model, called the littlest Higgs
model has only 3 free parameters:  the Higgs mass, the mass of the heavy photon and the mixing between standard and heavy quarks.
Because of electroweak constraints the spectrum is rather light, with $m_{A_H}<300$~GeV and $m_{T_+}<1$TeV. 
The production of heavy quarks 
which further decay into top quarks and the heavy photon is therefore quite large.
A determination of the mass of the $T_+$ quark as well as some combination of the Higgs, DM candidate and $T_-$ quark masses are
enough to overconstrain the model and make a "LHC prediction" of the DM relic abundance with a precision of 10\% even without
measuring precisely the mass of the heavy photon~\cite{Matsumoto:2009ki}.

\section{CONCLUSION}

Understanding the nature of dark matter is an exciting challenge for the LHC. 
While the prospects for discovering physics beyond the standard model at the TeV scale are excellent, a precise 
determination of the properties of the DM particle, to the level
where the theoretical predictions of DM observables reach the precision of the cosmological measurements 
is much more difficult. 

\end{document}